\documentclass[twocolumn,pre,superscriptaddress]{revtex4}
\usepackage{graphicx}
\usepackage{amsmath,amssymb}
\usepackage{tikz}
\usetikzlibrary{matrix}

\def\beqa{\begin{eqnarray}}
\def\eeqa{\end{eqnarray}}
\def\a={&=&}

\newcommand{\pa}[1]{\left( {#1} \right)}
\newcommand{\ffc}[2]{\includegraphics[scale=#1]{#2}}

\begin{document}
\title[]{
Assortative clustering in a one-dimensional population with
replication strategies
}
\author{Sunhee Chae}
\affiliation{Department of Physics and Astronomy, Sejong University, Seoul
05006, Korea}
\author{Nahyeon Lee}
\affiliation{Department of Physics and Astronomy, Sejong University, Seoul
05006, Korea}
\author{Seung Ki Baek}
\email[]{seungki@pknu.ac.kr}
\affiliation{Department of Physics, Pukyong National University, Busan 48513,
Korea}
\author{Hyeong-Chai Jeong}
\email[]{hcj@sejong.ac.kr}
\affiliation{Department of Physics and Astronomy, Sejong University, Seoul
05006, Korea}
\date{\today}
\begin{abstract}
In a geographically distributed population, assortative clustering plays an
important role in evolution by modifying local environments.
To examine its effects in a linear habitat,
we consider a one-dimensional grid of cells, where each cell is either empty or
occupied by an organism whose replication strategy is genetically inherited to
offspring.
The strategy determines whether to have offspring in surrounding
cells, as a function of the neighborhood configuration. If more than
one offspring compete for a cell, then they can be all exterminated due to the cost
of conflict depending on environmental conditions.
We find that the system is more densely
populated in an unfavorable environment than in a
favorable one because only the latter has to pay the cost of conflict.
This observation agrees reasonably well with a mean-field analysis which takes
assortative clustering of strategies into consideration.
Our finding suggests a possibility of intrinsic nonlinearity
between environmental conditions and population density when an evolutionary
process is involved.
\end{abstract}
\maketitle

\section{Introduction}
``Space exists so that everything doesn't happen to you,'' says Susan Sontag.
Spatiality often means being exempt from interacting with all others:
One may be surrounded by more favorable neighbors than the average
or the opposite, when the spatial configuration is nonuniform. If
the local environments experienced by individuals differ from place to place,
then it implies different selection pressure in terms of evolution, which
can shape the local environments even more differently. If such a feedback loop
forms, then individuals can break away from the evolutionary path that would have
been followed in a well-mixed population. For this reason, the roles of
spatiality in evolution have been studied extensively in the
literature~\cite{nakamaru1997evolution,hauert2004spatial,szabo2005phase,fu2010invasion,javarone2018host}.

To be more specific, let us consider a model of cellular automata,
one of the simplest models of life in spatial dimensions, yet with the
possibility of genuine complexity in its
behavior~\cite{gardner1970mathematical,gardner1971mathematical,bak1989self,silvertown1992cellular,alstrom1994self,rendell2002turing,bak2013nature}.
In a cellular-automata model, the space is divided into discrete cells, and the
cells can be occupied by ``organisms'' that replicate themselves according to
mechanistic laws. To put such a model into an evolutionary context,
we would like to point out the following:
The replication process would generate different copies with small errors in
practice, and each of the different copies would also have different
efficiency in replicating itself. In other words, they must be subject to
an evolutionary process of mutation and selection.

In this work, we will study evolution of such cellular organisms {\it in silico}
by assigning a replication strategy to each of them.
The strategy is transmitted genetically to offspring, and it has
to compete with others in neighboring cells. This defines a game in the sense of
game theory because an organism's payoff, identified with the number of
offspring, will depend on its neighbors' strategies as well as on its own.
An aggressive strategy would produce as many offspring as possible, invading the
territories of other strategies. Even if it incurs extra cost of conflict and
thus reduces the total size of the population, it should have a higher chance to
spread than nonaggressive ones.
If we regard the total population growth as the collective interest of life, then it
thus conflicts, at least partially, with individual interests of the selfish
genes that encode replication strategies.
However, a paradox of evolution is that self-interested behavior is not always
favored by selection~\cite{smith1982evolution,javarone2018statistical}, provided
that the dynamical rule permits assortative clustering of players who conform to
collective
interests~\cite{nowak2004evolutionary,fletcher2009simple,jeong2014optional,javarone2017evolutionary,bahk2019long}.
This study will show that such an assortative effect can be induced in a spatial
game by a simple mechanism, whereby defection from collective interests is
successfully suppressed. As a consequence, the mechanism introduces nonlinearity
in the relation between environmental conditions and population density.

This work is organized as follows: In the next section, we introduce our model.
The Monte Carlo simulation result will be presented in Sec.~\ref{sec:result}.
After explaining the observed behavior with a mean-field approximation in
Sec.~\ref{sec:discussion}, we conclude this work in Sec.~\ref{sec:summary}.

\section{Model}
Let us consider a group of organisms living on a one-dimensional grid with the
periodic-boundary conditions to see the assortative effect most clearly.
In ecology, such a one-dimensional structure describes a habitat constrained by
linear environmental features such as rivers or
shorelines~\cite{fisher1937wave,slaght2013home}, and it is also physically
relevant to studying dynamic processes in
$(1+1)$ dimensions~\cite{wolfram1984cellular,lavrentovich2013radial}.
Each grid cell is indexed by $x$, and its
occupancy is denoted by $n_x$: It can be either empty with $n_x=0$ or occupied
by one of the organisms with $n_x=1$.
Time $t$ is also a discrete variable, under the
assumption that the organisms have nonoverlapping generations. The model
consists of two parts, i.e., replication and mutation.

In the replication process, every organism produces
an offspring with the same strategy in its own cell. At the same time,
it may also produce offspring in neighboring cells.
Therefore, at the beginning of a new
generation, the number of offspring in a cell can sometimes be greater than
$1$. For example, let us imagine that only two neighboring cells, $x-1$ and $x$,
are occupied in an otherwise empty system by organisms with strategies $I$ and
$J$, respectively.
However, as implied by $n_x \le 1$, each cell can barely support a single adult:
If the $I$-player at $x-1$ produces two offspring $i$ and $i'$, one in its own
cell and the other in a neighboring cell $x$, then the latter will compete with the
$J$-player's offspring $j$ born in $x$. By assumption, they all die with
probability $1-\alpha$, leaving the cell $x$ empty, as a result of exhausting
competition. Here we have introduced a parameter $\alpha$ between $0$ and $1$,
which can be interpreted as the favorability of the environment. With
probability $\alpha$, the cell remains occupied, i.e., $n_x(t=2)=1$, in
which case we randomly choose one between $i'$ and $j$ as the survivor. If $i'$
is chosen, then it will have grown into an $I$-player at $t=2$; otherwise, we will
have a $J$-player in $x$ again. The above explanation can be schematically
represented as follows:
\begin{equation}
\vcenter{\hbox{
\begin{tikzpicture}
  \matrix (m) [matrix of math nodes,row sep=3em,column sep=4em,minimum width=2em]
  {
     t=2 & I & \Omega\\
     t=1 & I & J\\
     };
  \path[-stealth]
    (m-1-2.east|-m-1-3) edge [dashed,-] (m-1-3)
    (m-2-2.east|-m-2-3) edge [dashed,-] (m-2-3)
    (m-2-2) edge [] node [left] {$i$} (m-1-2)
    (m-2-2) edge [] node [left] {$i'$} (m-1-3)
    (m-2-3) edge [] node [right] {$j$} (m-1-3);
\end{tikzpicture}
}} \nonumber
\end{equation}
with
\begin{equation}
\hspace{1.5cm} \Omega = \left\{
\begin{array}{ll}
E & \text{with prob.~}1-\alpha\\
I & \text{with prob.~}\alpha/2\\
J & \text{with prob.~}\alpha/2,
\end{array}
\right.
\label{eq:n2}
\end{equation}
where $E$ denotes that the cell is empty.
Similarly, let us imagine that the system starts with only three organisms,
which
occupy three consecutive cells, $x-1$, $x$, and $x+1$, and play strategies $I$,
$J$, and $K$, respectively. If the $I$- and $K$-players produce their offspring
in the $J$-player's cell $x$, then the competition of the three will be more intense
than the above case of two competitors. We describe this situation by
assuming that the focal cell $x$ becomes empty with probability $1-\alpha^2$,
which is greater than $1-\alpha$ for $\alpha \in (0,1)$. If the cell remains
occupied, i.e., $n_x(t=2)=1$, one of the three competitors is chosen randomly as
the survivor. This example can thus be represented as follows:
\begin{equation}
\vcenter{\hbox{
\begin{tikzpicture}
  \matrix (m) [matrix of math nodes,row sep=3em,column sep=4em,minimum width=2em]
  {
     t=2 & I & \Omega & K\\
     t=1 & I & J & K\\};
  \path[-stealth]
    (m-1-2.east|-m-1-3) edge [dashed,-] (m-1-3)
    (m-1-3.east|-m-1-4) edge [dashed,-] (m-1-4)
    (m-2-2.east|-m-2-3) edge [dashed,-] (m-2-3)
    (m-2-3.east|-m-2-4) edge [dashed,-] (m-2-4)
    (m-2-2) edge [] node [left] {$i$} (m-1-2)
    (m-2-2) edge [] node [left] {$i'$} (m-1-3)
    (m-2-3) edge [] node [right] {$j$} (m-1-3)
    (m-2-4) edge [] node [right] {$k$} (m-1-4)
    (m-2-4) edge [] node [right] {$k'$} (m-1-3);
\end{tikzpicture}
}} \nonumber
\end{equation}
with
\begin{equation}
\hspace{0.5cm} \Omega = \left\{
\begin{array}{ll}
E & \text{with prob.~}1-\alpha^2\\
I & \text{with prob.~}\alpha^2/3\\
J & \text{with prob.~}\alpha^2/3\\
K & \text{with prob.~}\alpha^2/3.
\end{array}
\right.
\label{eq:n3}
\end{equation}

When an organism exists in a cell $x$, we assume that its replication strategy
takes into account $n_{x-1}$ and $n_{x+1}$, that is, the occupancy of
neighboring cells.
We thus have to distinguish four cases, denoted by
$\nu_x\equiv 2^1 \times n_{x-1} + 2^0 \times n_{x+1}$,
so that $\nu_x$ can take a value from $\{0, 1, 2, 3\}$.
This variable can be conveniently represented in binary:
If $n_{x-1}= n_{x+1} =1$, for example, then we can write $\nu_x = 11$.
Let $b_{x \to y}$ be a binary variable for the replication behavior
which represents whether the organism in $x$ produces an offspring in
$y$:
If it does, then $b_{x\to y}=1$, and $0$ otherwise.
Note that $b_{x\to x}=1$ because the organism will always produce an
offspring in its own cell.
Then the strategy of the organism in a cell $x$ is determined by
its replication behavior
$\beta_x \equiv 2^1 \times b_{x\to x-1} + 2^0 \times b_{x\to x+1}$
as a function of $\nu_x \in \{00, 01, 10, 11\}$.
The replication behavior
$\beta_x$ can also be represented in binary,
for example, $\beta_{x} (00) = 11$
if $b_{x\to x-1} = b_{x\to x+1}=1$ for $\nu_x = 00$.
It means that the strategy will produce offspring in both the
neighboring cells when they are empty.
We now represent the strategy as an eight-digit binary number
by arranging $\beta_x(\nu_x)$ in descending order of
$\nu_x$ from 11 to 00.
The most aggressive strategy will always produce offspring in the
neighboring cells by assigning $\beta_x = 11$ to all four $\nu_x$'s.
This strategy can thus be indexed as $11111111$ in binary, which
corresponds to $255$ in decimal. Note that the subscript $x$ can actually
be dropped in the above description because the strategy itself has no
dependence on the position. As another example, the most inactive strategy
should have $\beta = 00$ for every $\nu \in \{0,1,2,3\}$, hence $00000000=0$ as
its index, because it will never invade the neighboring cells. Among $256$
possible strategies between these two extremes, Table~\ref{table:27} shows a
nontrivial strategy that produces offspring only in empty neighboring cells,
whose index is calculated as $27$ in decimal. As will be shown by numerical
simulation below, this turns out to be one of the most important strategies in
our model.

\begin{table}
\begin{tabular}{l|cccc}\hline\hline
neighboring-cell occupancy $\nu$ & $11$ & $10$ & $01$
& $00$\\%\hline
replication behavior $\beta$ for each $\nu$ & $00$ & $01$ & $10$ &
$11$\\\hline
%contribution in decimal & $0$ & $16$ & $8$ & $3$\\\hline
strategy index & \multicolumn{4}{r}{$27$}\\\hline\hline
\end{tabular}
\caption{Example of a replication strategy indexed as $27$, which
produces offspring only in empty neighboring cells. Such behavior is
characterized by $\beta = \text{NOT}~\nu$, where NOT means logical negation on
each bit.
As shown in the first line, we sort $\nu$ in descending order from 11 to
00, so the binary representation of this strategy is obtained as
$00011011$ (the second row), which corresponds to $27$ in decimal.
}
\label{table:27}
\end{table}

In the presence of environmental noise, the strategic information may be lost
in the course of replication. Thus, we assume that an offspring's strategy may
change to an arbitrary one in the set of available strategies $\mathcal{S}
\equiv \{ 0, 1, \ldots, 255\}$ with small mutation probability $\mu \ll 1$.
The mutation process is also important from a computational point of view:
We will calculate time-averaged quantities from a Monte Carlo method.
This would not be justified without mutation because the system might cease to
be ergodic when it reaches an absorbing state consisting of a single strategy.

%%%%%%%%%%%[Fig. 1]%%%%%%%%%%%%%%%%
\begin{figure}[t!]
\ffc{.33}{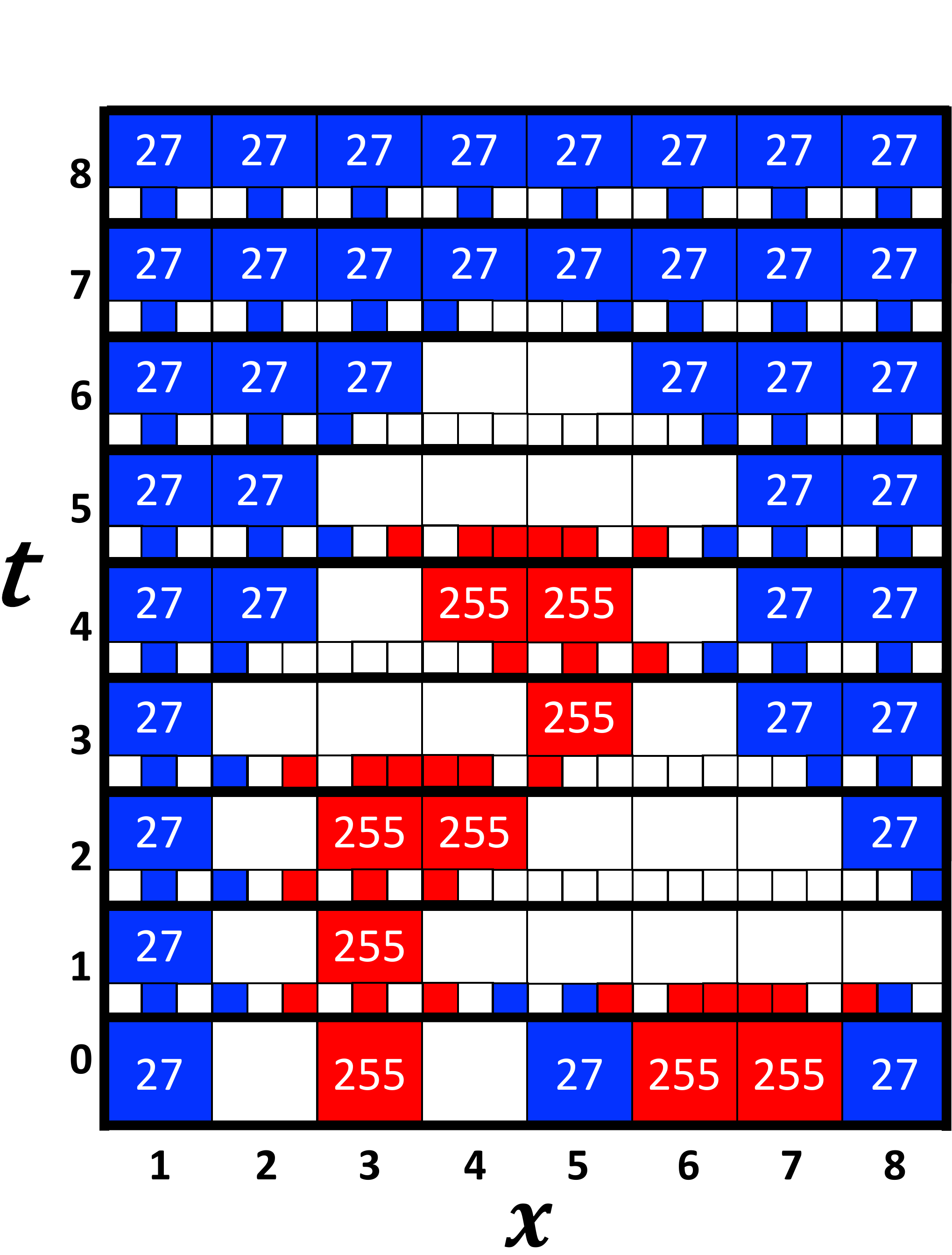}
\caption{
Evolution of a population which is initially composed of strategies
$27$ (blue) and $255$ (red). The horizontal axis represents the
spatial dimension under the periodic boundary conditions, and the vertical axis
represents time in units of generations.
The initial configuration is given at the bottom ($t=0$). Between generations,
we have draw three small blocks for each cell to represent the offspring
produced in that cell. The environment is assumed to be extremely unfavorable
($\alpha=0$), so the cell will become blank if more than one offspring are
produced there.
}
\label{f1}
\end{figure}

Figure~\ref{f1} illustrates our model by showing how a population of the
above-mentioned two strategies, i.e., $27$ and $255$, evolves on a
one-dimensional ring with $L=8$ cells.
Both the environmental parameter $\alpha$ and the mutation probability $\mu$ are
set to be zero to help follow the rules in a fully deterministic way.
Note that all the cells are updated in parallel as $t$ increases
by one in this example, and this will also be the case of our Monte Carlo
calculation in the next section (see Ref.~\onlinecite{saif2009prisoner} for
possible effects of update rules on time evolution). However, the long-time
behavior presented below shows no significant difference even when we
use a random asynchronous update rule.

\section{Result}
\label{sec:result}

%%%%%%%%%%%[Fig. 2]%%%%%%%%%%%%%%%%
\begin{figure}[t!]
\ffc{.13}{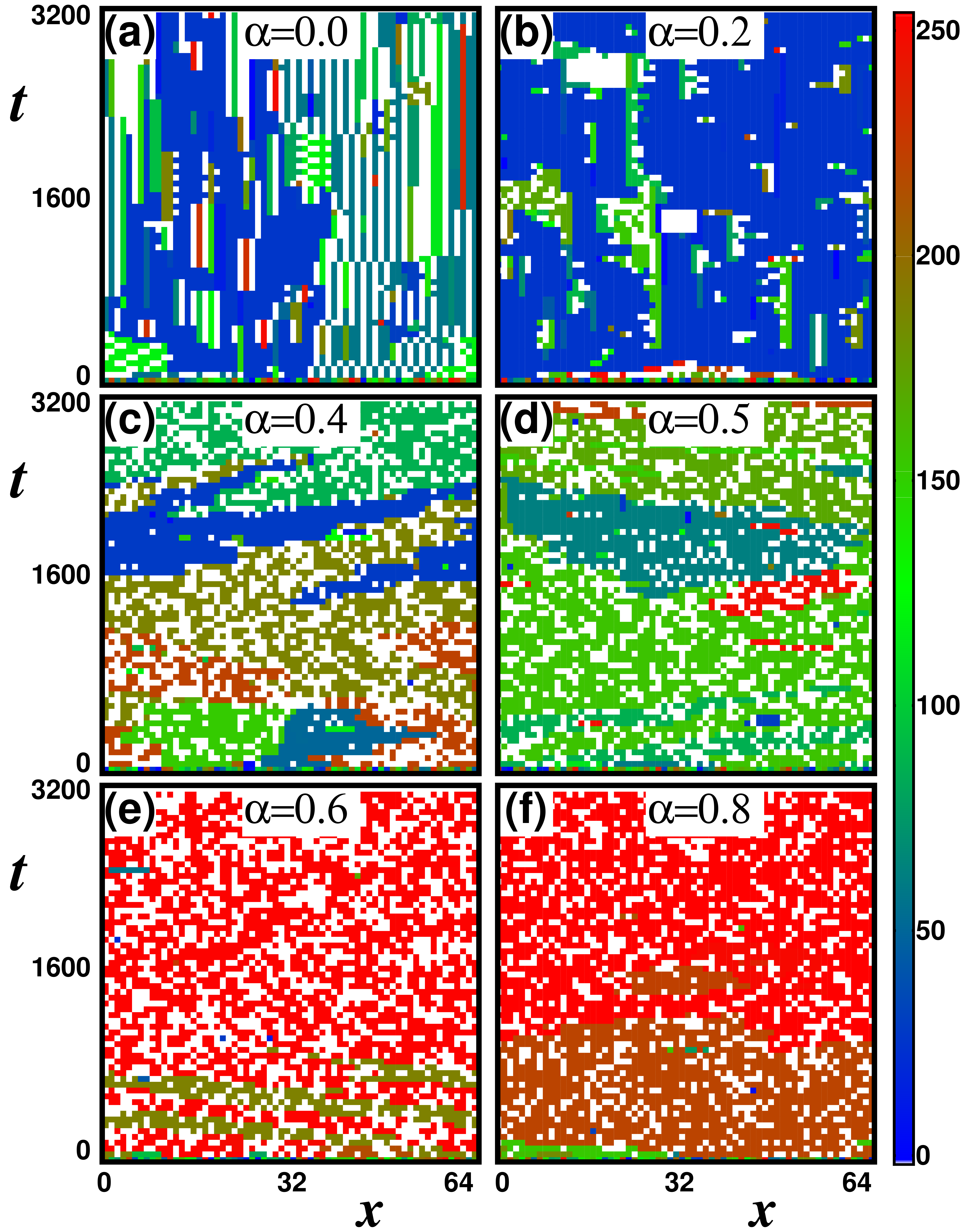}
\caption{
Effects of the environmental favorability $\alpha$. Each panel shows a
simulation result with a different value for $\alpha$, ranging from $0.0$ to
$0.8$. The mutation probability is fixed at $\mu=10^{-3}$.
As in Fig.~\ref{f1}, the horizontal axis represents the spatial
dimension. The vertical axis represents time, along which we have sampled the
data at every $50$ generations. Initially at $t=0$, every cell is occupied by an
organism with a random strategy drawn from $\mathcal{S}$. The colors represent
strategy indices from $0$ to $255$ (see the color box on the right), and the
white cells are empty.
}
\label{f2-config}
\end{figure}

Let us now include all the 256 strategies of $\mathcal{S}$ and simulate the
model on a larger ring structure with $L=64$ cells (Fig.~\ref{f2-config}).
Initially at $t=0$, every cell is occupied by an organism with a randomly drawn
strategy from $\mathcal{S}$. The colors represent strategy indices from $0$ to
$255$. Bluish strategies do not produce offspring in neighboring cells when they
are occupied. In other words, they are characterized by $\beta (\nu=11) = 00$.
On the other hand, reddish strategies aggressively produce offspring in such a
situation by having $\beta(\nu=11) = 11$. Greenish strategies are in between, so
they have either $\beta(\nu=11) = 01$ or $10$.

Figure~\ref{f2-config} shows which class of strategies are favored depending on
$\alpha$: When $\alpha = 0.0$ or $0.2$, the system is bluish, and the reason is
that aggressive strategies are very likely to be removed with such a low value
of $\alpha$. The bluish cluster is usually dense because these strategies tend
to avoid conflict with neighbors. On the other hand, reddish strategies take
over when $\alpha = 0.6$ or $0.8$, but their cluster is porous, and the porosity
will gradually vanish as $\alpha \to 1$.

%%%%%%%%%%%[Fig. 3]%%%%%%%%%%%%%%%%
\begin{figure}[t!]
\ffc{.65}{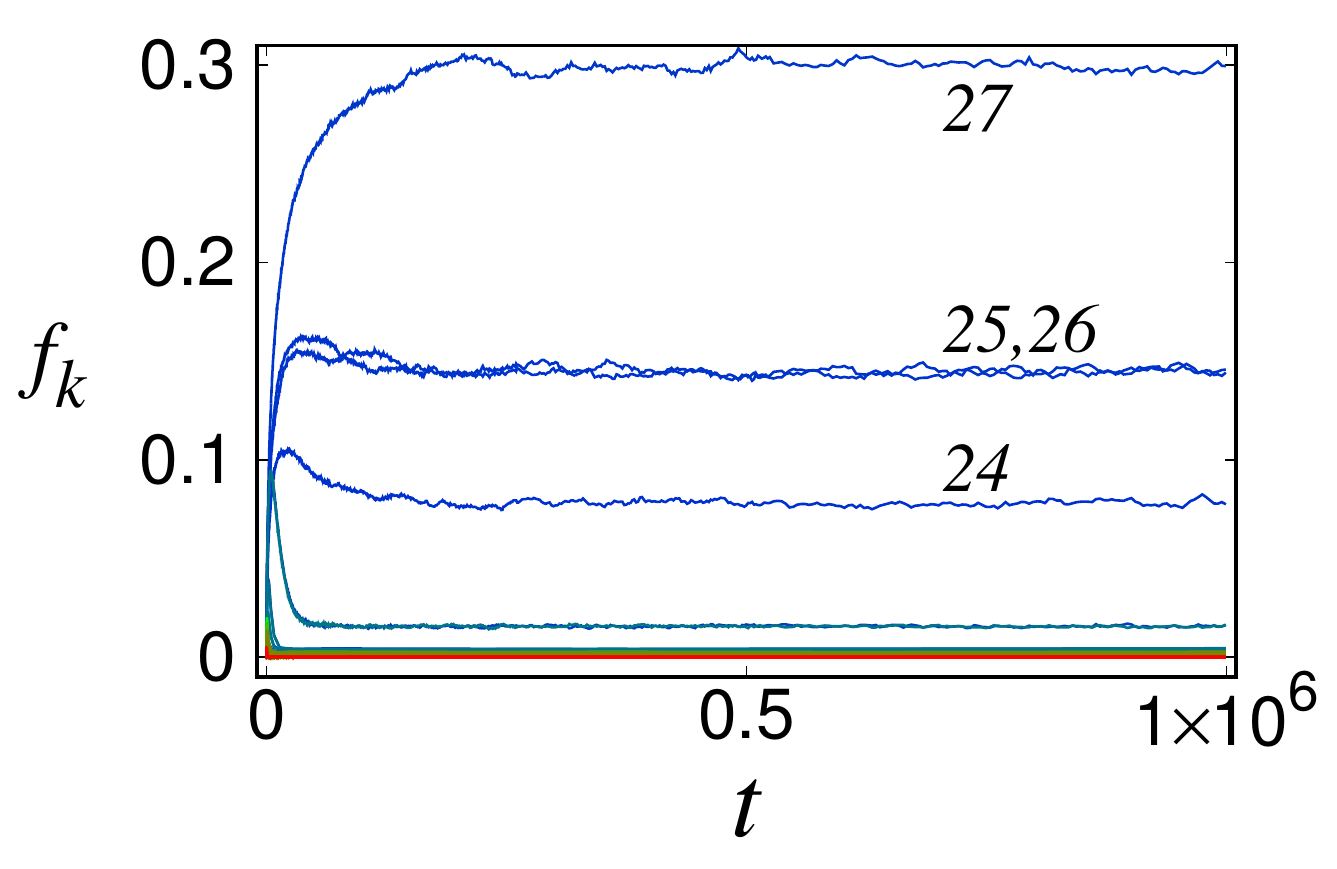}
\caption{
Frequencies of strategies under $\alpha=0.1$.
The system size is $L=1024$, the mutation rate is $\mu=10^{-3}$, and all the
results are averaged over $M=10^3$ independent realizations.
The colors represent strategy indices from $0$ to $255$ as in
Fig.~\ref{f2-config}. Initially at $t=0$, every strategy starts with an equal
frequency, but selection favors strategy $27$ and its variants.
}
\label{f3-t-frq}
\end{figure}

To quantify the behavior, we calculate the ensemble--averaged frequency of
strategy $k$ as follows:
\begin{equation}
f_k (t) = \frac{1}{M} \sum_{m=1}^M \frac{N_k^{(m)}(t)}{L},
\label{eq:f_k}
\end{equation}
where $M$ is the number of independent Monte Carlo realizations
and $N_k^{(m)} (t)$ is the number of organisms playing strategy $k$ at time
$t$ in the $m$th realization (Fig.~\ref{f3-t-frq}). To obtain its value in a
steady state, we remove transient behavior for a certain initial period $T$ and
then take an average over $P$ generations:
\begin{equation}
\phi_k = \frac{1}{P} \sum_{t=T+1}^{T+P} f_k (t).
\label{eq:phi_k}
\end{equation}
The total density of population,
\begin{equation}
\rho = \sum_{k \in \mathcal{S}} \phi_k,
\label{eq:rho}
\end{equation}
is a measure of collective interests for this group of organisms.

%%%%%%%%%%%[Fig. 4]%%%%%%%%%%%%%%%%
\begin{figure}[t!]
\ffc{.6}{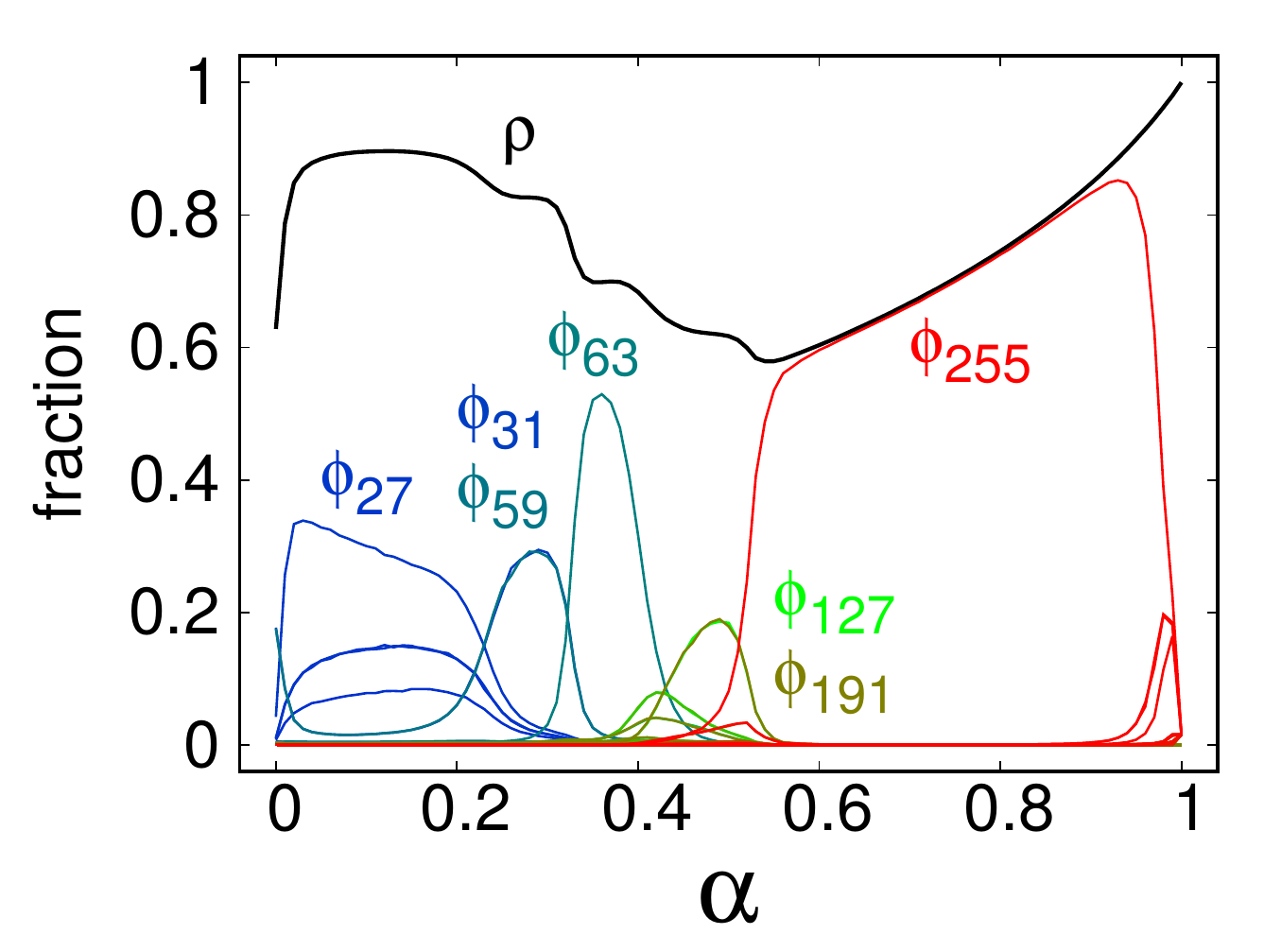}
\caption{Steady-state frequency of each strategy $k$ [Eq.~\eqref{eq:phi_k}] as a
function of $\alpha$. The colors of $\phi_k$'s are as given in
Fig.~\ref{f2-config}. The total density of the population [Eq.~\eqref{eq:rho}]
is represented by the thick black line. We use the same $L$, $M$, and $\mu$ as
in Fig.~\ref{f3-t-frq}. The time average has been taken over $P=10^5$
generations, after removing transients for the first $T=9 \times 10^5$
generations.
}
\label{f4-s-frq}
\end{figure}

Let us check how these observables behave as $\alpha$ varies.
Figure~\ref{f4-s-frq} shows that an unfavorable environment with small $\alpha$
tends to favor bluish nonaggressive strategies such as $27$, and they are
replaced by more and more aggressive ones as $\alpha$ increases, which is
entirely consistent with Fig.~\ref{f2-config}.
Note that strategies $31$ and $59$ exhibit identical behavior because they are
related by left-right symmetry, and the same statement holds between $127$
and $191$. An interesting point is that the total density of the population
\emph{decreases} as the environment becomes more and more favorable between
$0.2$ and $0.5$.

\section{Discussion}
\label{sec:discussion}

%%%%%%%%%%%[Fig. 5]%%%%%%%%%%%%%%%%
\begin{figure}[t!]
\ffc{.29}{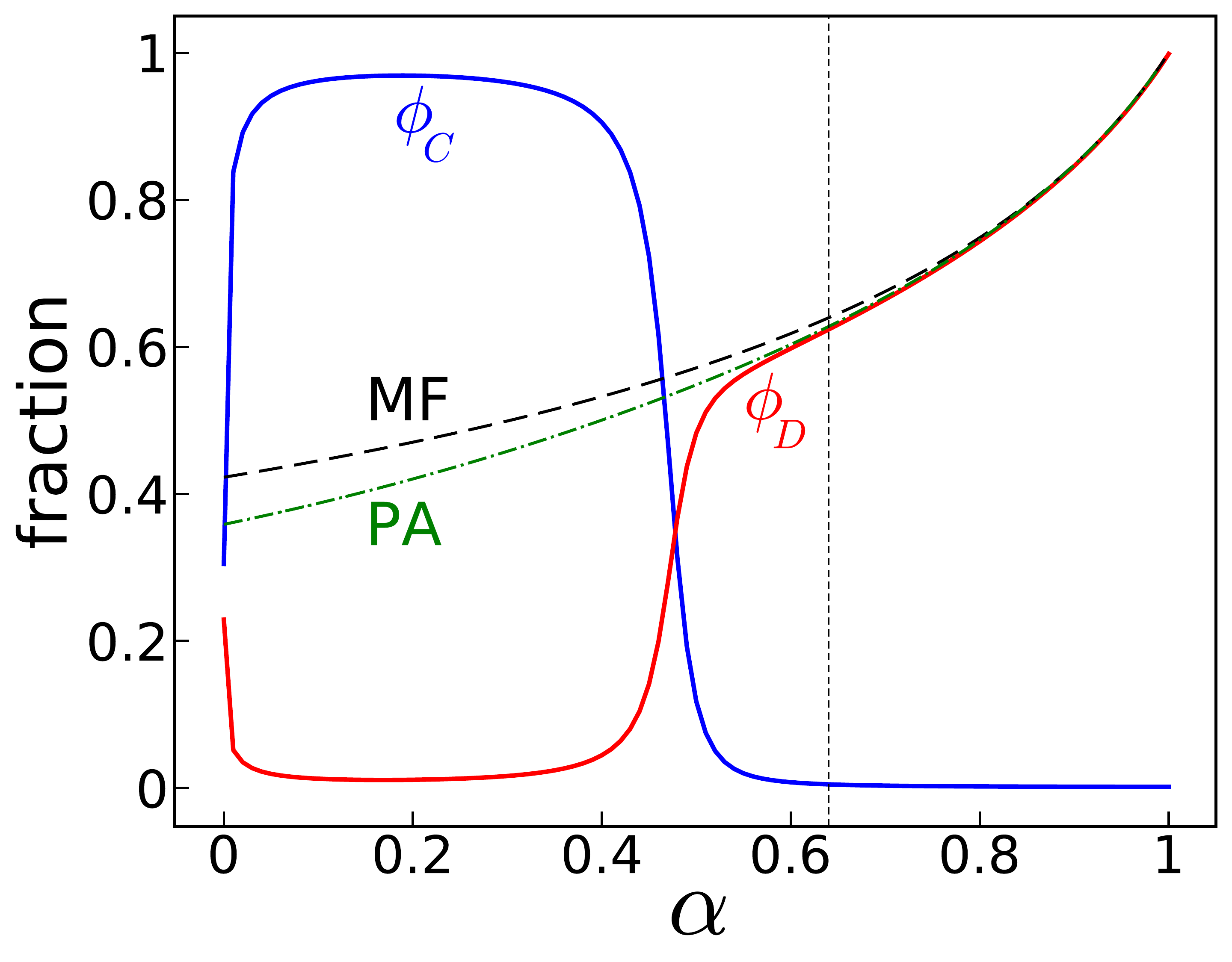}
\caption{
  Simplified dynamics with a reduced set of strategies, i.e., $\{ C=27, D=255 \}$.
  The other simulation parameters are the same as in Fig.~\ref{f4-s-frq}.
  The $\phi_C$ and $\phi_D$ curves represent steady-state frequencies
  of $C$ and $D$, whereas the dashed and dash-dotted lines mean
  the simple mean-field (MF) and pair approximation (PA) results, respectively,
  for a $(D+E)$ cluster.
  The vertical dotted line is an approximate value of $\alpha^\ast$,
  obtained by solving Eq.~\eqref{eq:interface}, above which $D$
  invades $C$.
}
\label{f5-two}
\end{figure}

To illustrate the basic picture, it is instructive to work with a reduced set of
strategies. We choose $27$ and $255$, the most favored ones for small and large
$\alpha$'s, respectively (Fig.~\ref{f4-s-frq}). The former strategy is able to
increase the population density $\rho$ up to $100\%$ by replicating itself in a
nonaggressive way. Thus, it may be called a ``cooperating'' strategy. The latter
strategy is the most aggressive one, and we may call it a ``defecting'' strategy.

Let us consider three consecutive cells, each of which is either
$C$ (cooperating), $D$ (defecting), or $E$ (empty).
From the configuration of these three cells, we can discuss the
replication dynamics in the middle cell.
Let $\eta_{_{C}}^{+}$ be the rate for an empty cell to be occupied
by a cooperator.
By enumerating all the possible cases, we see that
\begin{eqnarray}
\!\!\eta_{_C}^{+}
\!&\!=\!&\! \left(\phi_{_{EEC}}\!+\!\phi_{_{CEE}} \right)
+ \pa{\phi_{_{CED}}\!+\!\phi_{_{DEC}}} \frac{\alpha}{2}
+ \phi_{_{CEC}} \alpha,
\end{eqnarray}
where $\phi_{_{XYZ}}$ is the frequency for the three consecutive cells to have
states $X$, $Y$, and $Z$, respectively.
On the other hand,
a $C$ cell becomes either a $D$ cell or an $E$ cell
with a rate of
\begin{eqnarray}
\eta_{_{C}}^{-}
&=& \pa{\phi_{_{ECD}}+ \phi_{_{CCD}} + \phi_{_{DCE}} + \phi_{_{DCC}}}
\pa{1-\frac{\alpha}{2}} \nonumber \\
&&\!\!+\ \phi_{_{DCD}} \pa{1-\frac{\alpha^2}{3}}.
\end{eqnarray}
Similarly, we can write the rates
$\eta_{_D}^\pm$ for creation or annihilation of $D$ cells.
However, to know $\phi_{_{XYZ}}$,
the statistics of \emph{five} consecutive cells is required,
and this hierarchy generally goes on
{\it ad infinitum}~\cite{krapivsky2010kinetic,yi2015interrupted}.
As an approximation, we will factorize
$\phi_{_{XYZ}}$ into $\phi_{_{X}} \phi_{_{Y}} \phi_{_{Z}}$,
where $\phi_{_{X}}$ denotes the frequency of
$X$ cells [see Eq.~\eqref{eq:phi_k}],
and this mean-field approximation is
valid in the absence of spatial correlations.
Figure~\ref{f2-config}(f)
suggests that $D$ cells form a homogeneous
mixture with $E$ cells
(without strong spatial correlations)
for large $\alpha$ in the steady states.
We can estimate the frequency of $D$ cells
using a mean-field approximation in such $(D+E)$ clusters.
In these clusters,
$\eta_{_D}^\pm$
is given by
\begin{eqnarray}
  \eta_{_{D}}^{+}
  &=& \pa{\phi_{_{EED}} + \phi_{_{DEE}}} + \phi_{_{DED}} \alpha,\\
  \eta_{_{D}}^{-}
  &=& \pa{\phi_{_{EDD}} + \phi_{_{DDE}}}\pa{1-\alpha} + \phi_{_{DDD}} \pa{1-\alpha^2}.
\end{eqnarray}
Equating $\eta_{_{D}}^{+}$ and $\eta_{_{D}}^{-}$,
the equilibrium frequency of
$D$ cells in a $(D+E)$ cluster is obtained as
\begin{eqnarray}
  \phi_{_{D}}
  &=& \frac{6-3\alpha - \sqrt{12-12\alpha +\alpha^2}}{2\pa{3-3\alpha +\alpha^2}}
      \label{eq:f_D}\\
  &=& 1 + 2\epsilon + 7\epsilon^2 + 35 \epsilon^3 + \ldots,
      \label{eq:taylor}
\end{eqnarray}
where we used $\phi_{_{E}} = 1-\phi_{_{D}}$
and set $\epsilon \equiv \alpha-1 < 0$ (Fig.~\ref{f5-two}).
Equation~\eqref{eq:f_D} agrees well with
our numerical results for large $\alpha$, implying that the whole system can be
described as a single $(D+E)$ cluster. On the other hand, the system is mostly
filled with $C$ cells for small $\alpha$. This suggests existence of a
transition between $C$ and $(D+E)$ phases at a certain threshold
$\alpha = \alpha^\ast$.

To estimate $\alpha^\ast$,
let us assume that a $(D+E)$ cluster has an interface with a $C$ cluster.
The most probable situation for growth of the $C$ cluster is found when
the two nearest cells to the interface on the $(D+E)$ side are empty.
The simplest estimate for this probability would be $\phi_{_{E}}^2 =
(1-\phi_{_{D}})^2$, under the assumption that the bulk behavior inside a
$(D+E)$ cluster is mostly valid even in the vicinity of the interface.
The second contribution is given by another configuration in
which $D$ and $C$ compete for an empty cell in the middle, and this
contributes $\phi_{_{D}} \left( 1-\phi_{_{D}} \right) (\alpha/2)$ because $C$
wins with probability $\alpha/2$. On the other
hand, the $(D+E)$ cluster can proceed by one cell with probability
$\phi_{_{D}} \times (\alpha/2)$ because the front must be filled with $D$ and
the invasion succeeds with probability $\alpha/2$. If we compare these two
events, then the latter becomes more probable for large $\alpha$, and the threshold
value is estimated by equating them, i.e.,
\begin{equation}
(1-\phi_{_{D}})^2 + \phi_{_{D}} \left( 1-\phi_{_{D}} \right) (\alpha/2) =
\phi_{_{D}} \times (\alpha/2),
\label{eq:interface}
\end{equation}
which, together with Eq.~\eqref{eq:f_D}, results in $\alpha^\ast \approx 0.64$
(Fig.~\ref{f5-two}). We note that this may well be an overestimate because
the actual frequency of $D$ is likely to be higher than
predicted by Eq.~\eqref{eq:f_D} near the interface, where the competition
between $C$ and $D$ would be less intense than between two $D$'s.

The above mean-field calculation can be modified by using the pair
approximation~\cite{joo2004pair}, according to which three-point and
four-point correlation functions are approximated as
\begin{equation}
\phi_{_{XYZ}} \approx \frac{\phi_{_{XY}} \phi_{_{YZ}}}{\phi_{_{Y}}}
\label{eq:pa3}
\end{equation}
and
\begin{equation}
\phi_{_{XYZW}} \approx \frac{\phi_{_{XY}} \phi_{_{YZ}}
\phi_{_{ZW}}}{\phi_{_{Y}} \phi_{_{Z}}},
\label{eq:pa4}
\end{equation}
respectively (see
Refs.~\onlinecite{dickman1988mean,ben1992mean,mendoncca2011extinction}
for further modification beyond the pair approximation).
If we deal with a $(D+E)$ cluster, then we need five correlation functions, i.e.,
$\phi_{_{D}}$,
$\phi_{_{E}}$,
$\phi_{_{DD}}$,
$\phi_{_{DE}}=\phi_{_{ED}}$, and
$\phi_{_{EE}}$,
but only two of them are independent because
$\phi_{_{D}} = 1-\phi_{_{E}} =
\phi_{_{DD}} + \phi_{_{DE}} = 1 - \left(
\phi_{_{EE}} + \phi_{_{ED}} \right)$.
If we find $\phi_{_{D}}$ and $\phi_{_{DD}}$, for example, then the other three
are determined by these relations.
Regarding $\phi_{_{DD}}$,
the rates of creating and annihilating $DD$ cells are given as
\begin{eqnarray}
\eta_{_{DD}}^+ &=&
\left( \phi_{_{EDEE}} + \phi_{_{DEED}} + \phi_{_{EEDE}} \right)\nonumber\\
&&+ \alpha
\left( \phi_{_{EDED}} + \phi_{_{DDEE}} + \phi_{_{EEDD}} + \phi_{_{DEDE}}
\right)\nonumber\\
&&+ \alpha^2 \left( \phi_{_{DDED}} + \phi_{_{DEDD}} \right)
\end{eqnarray}
and
\begin{eqnarray}
\eta_{_{DD}}^- &=&
\left(1 - \alpha^4 \right) \phi_{_{DDDD}}
+\left(1 - \alpha^3 \right) \left( \phi_{_{EDDD}} + \phi_{_{DDDE}}
\right)\nonumber\\
&& +\left(1 - \alpha^2 \right) \phi_{_{EDDE}},
\end{eqnarray}
respectively. By solving $\eta_{_{D}}^+ = \eta_{_{D}}^-$ and
$\eta_{_{DD}}^+ = \eta_{_{DD}}^-$ with the pair approximation
[Eqs.~\eqref{eq:pa3} and \eqref{eq:pa4}],
we obtain $\phi_{_{D}}$
as a function of $\alpha$~\cite{Mathematica}
which is shown as the dash-dotted curve in Fig.~\ref{f5-two}.
Although its explicit expression is not illuminating, a few points are worth
mentioning: First, the pair-approximated version of $\phi_{_{D}}$ has the same
Taylor series to the order of $\epsilon^3$ as given by the mean-field
calculation [see Eq.~\eqref{eq:taylor}].
Second, if we also write $\phi_{_{DD}}$ as a function of $\alpha$, then
we obtain the \emph{connected} correlation function $\tilde{\phi}_{_{DD}} \equiv
\phi_{_{DD}} - \phi_{_{D}}^2 = -4\epsilon^3 + \ldots$, which is indeed small
and thus consistent with the mean-field-like ideas behind our approximate
calculation. Third, the system has four solution branches, and the physical
solution, having both $\phi_{_{D}}$ and $\phi_{_{DD}}$ inside the unit interval,
changes its branch at $\alpha \approx 0.60$, which might indicate an improved
estimate of $\alpha^\ast$.

%%%%%%%%%%%[Fig. 6]%%%%%%%%%%%%%%%%
\begin{figure}[t!]
\ffc{.068}{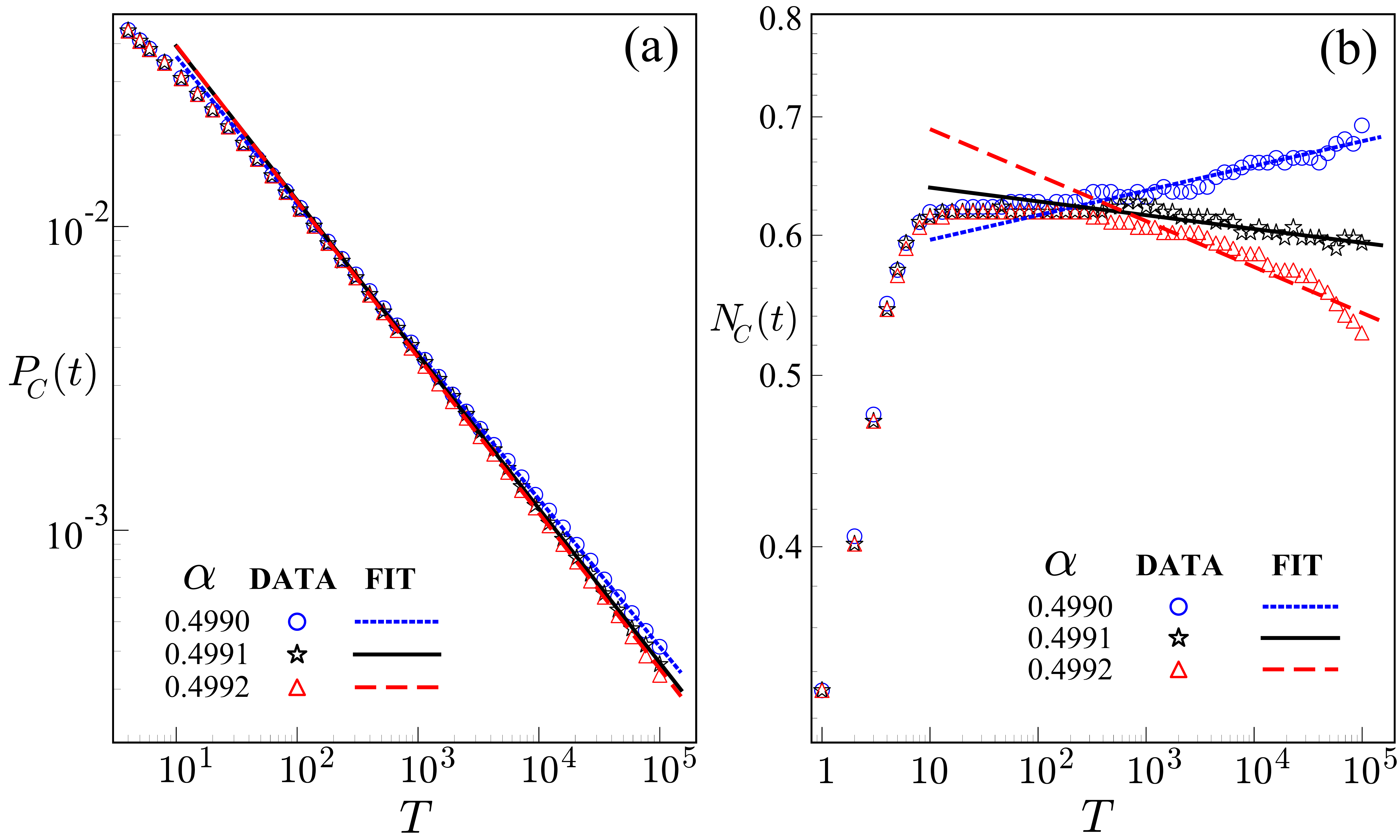}
\caption{
  (a) Survival probability of a single $C$-cell $P_{_C}(t)$ at time
  $t$ for $\alpha=0.4990$, $0.4991$, and $0.4992$, when all the other
  cells were initially filled with $D$.
  We observe power-law behavior $P_{_C}(t) \sim t^{-\delta}$ with
  $\delta = 0.50 \pm 0.02$.
  (b) The number of $C$ cells $N_{_C}(t)$ at time $t$ from the same
  initial configuration.
  For $\alpha = 0.4991 \pm 0.0001$ as in (a),
  they are described by $N_{_C}(t) \sim t^\theta$ with
  $\theta=-0.01 \pm 0.03$ as seen from the three straight lines.
  We have generated $4\times 10^5$ independent samples for each $\alpha$.
}
\label{fig:crit}
\end{figure}

\def\comment#1{}
\comment{
non-linear least squares to fit a function, f, to data
using scipy.
0.499
d= -0.4858701071077402
a= 0.11112129334254621
var_d= 3.4105268933033187e-09
var_a= 2.586294821320423e-07
0.4991
d= -0.5074968494768668
a= 0.12627903746500826
var_d= 1.137604911574579e-08
var_a= 8.568082102969428e-07
0.4992
d= -0.5132123875198508
a= 0.12861089166220463
var_d= 1.0051034423580848e-08
var_a= 7.583961119617394e-07
===============
0.499
d= 0.013938958305754957
a= 0.5775949655555447
var_d= 6.919387437678871e-09
var_a= 7.093855716333748e-07
0.4991
d= -0.007797214800565327
a= 0.649960350609817
var_d= 5.424416560367989e-09
var_a= 5.561189998019663e-07
0.4992
d= -0.02597308814163146
a= 0.7313346803524385
var_d= 3.1754878969150206e-08
var_a= 3.2555559194996407e-06
where the error mainly originates from the uncertainty in
$\alpha^\ast$.
}

To understand the nonequilibrium phase transition between $C$ and $D$ more
precisely~\cite{hinrichsen2000non,marro2005nonequilibrium,szabo2002phase}, we
conduct Monte Carlo simulation and observe the following quantity: Let
$P_{_C}(t)$ be the probability to have at least one
$C$ cell at time $t$ when the simulation started at $t=0$
with a single $C$ cell in a system filled with $D$.
The result in Fig.~\ref{fig:crit}(a) shows that it
decays as $P_{_C}(t) \sim t^{-\delta}$ at $\alpha^\ast = 0.4991 \pm 0.0001$.
From $4\times 10^5$ samples for each $\alpha$, we estimate the decay
exponent as $\delta = 0.50 \pm 0.02$, where the error mainly originates from
the uncertainty in $\alpha^\ast$.
The number of $C$ cells is another quantity expected to show
power-law behavior $N_{_C}(t) \sim t^\theta$ [Fig.~\ref{fig:crit}(b)],
and we estimate the exponent
as $\theta = -0.01 \pm 0.03$.
We have also obtained consistent results by exchanging
$C$ and $D$ in the initial configuration (not shown).

To conclude, our approximate calculation predicts that $C$ will densely occupy
the whole system if $\alpha < \alpha^\ast$. Otherwise, the system will be
occupied by a mixture of $D$ and $E$, among which the fraction of $D$ is
described by Eq.~\eqref{eq:f_D}. The total density of population should decrease
as $\alpha$ exceeds $\alpha^\ast$ because $\phi_{_{D}}(\alpha = \alpha^\ast)$ is
far smaller than $100\%$.
Our numerical results suggest that the behavior at $\alpha=\alpha^\ast$ can be
described by random walks of domain walls because
the survival probability behaves as
$P_{_C}(t) \sim t^{-\delta}$ with $\delta \approx 1/2$ and
the average number of $C$ cells is approximately constant.

\section{Summary}
\label{sec:summary}

To summarize, we have studied an evolutionary game in
which replication strategies are inherited by the next generation and the
survival probability in competition depends on neighbors' strategies as well as
one's own.
We have examined evolution of the population with varying the
environmental favorability that determines the chance of surviving competition.
Our finding is that the population sometimes flourishes better when
the survival probability is smaller because it eventually evolves to a more
cooperative strategy.
Although we have focused on a one-dimensional system to see the effects of
spatiality most clearly, it is entirely plausible that the effects will diminish
in higher dimensions and disappear in a well-mixed population. Exact identification of the critical dimension is left as a future work.

A common assumption in microeconomics is that production functions monotonically
increase in all inputs so that output quantities do not decrease when any
input quantity is increased. Our result suggests that the monotonicity
assumption may not always hold when an evolutionary process is involved, if we
regard $\alpha$ as a measure of input resources and the population density
$\rho$ as the output.
If the organisms under consideration are coupled with the input
resources through a predator-prey interaction, then it implies that the
coupling will be described as nonlinear, as opposed to the linear coupling
in the Lotka-Volterra type, due to the intraspecific interaction among
different behavioral strategies.
More specifically, the mean-field analysis discussed above
shows that assortative clustering can result in nonmonotonic behavior
through interfacial dynamics between two competing clusters.
It demonstrates the role of assortative clustering in evolution of cooperation
under the condition of resource scarcity.

\section*{Acknowledgments}
S.K.B. was supported by Basic Science Research Program through the
National Research Foundation of Korea (NRF) funded by the Ministry of
Education (Grant No. NRF-2020R1I1A2071670).
H.C.J. was supported by Basic Science Research Program through the
National Research Foundation of Korea (NRF) funded by the Ministry of
Education (Grant No. NRF-2018R1D1A1A02086101).

%\bibliography{google}
%\bibliographystyle{apsrev4-1}

\end{document}